\documentclass[11pt,a4paper]{article}
\usepackage{jheppub,multirow}

\title{\boldmath Interference effects for $H\to WW/ZZ\to\ell\bar{\nu}_\ell\bar{\ell}\nu_\ell$ searches in gluon fusion at the LHC}
\author{Nikolas Kauer}
\affiliation{Department of Physics, Royal Holloway, University of London, Egham Hill, Egham TW20 0EX, U.K.}
\emailAdd{n.kauer@rhul.ac.uk}

\abstract{$WW/ZZ$ interference for Higgs signal and continuum background as well 
as signal-background interference is studied 
for same-flavour $\ell\bar{\nu}_\ell\bar{\ell}\nu_\ell$ final 
states produced in gluon-gluon scattering at the LHC for light and heavy 
Higgs masses with minimal and realistic experimental selection cuts.
For the signal cross section, we find 
$WW/ZZ$ interference effects of ${\cal O}(5\%)$ at $M_H=126$ GeV.  
For $M_H\ge 200$ GeV, we find that $WW/ZZ$ interference is negligible.  
For the $gg$ continuum background, we also find that $WW/ZZ$ interference 
is negligible.  
As general rule, we conclude that non-negligible $WW/ZZ$ interference 
effects occur only if at least one weak boson of the pair is dominantly 
off-shell due to kinematic constraints.
The subdominant weak boson pair contribution induces a 
correction to the signal-background interference, which is 
at the few percentage point level before search selection cuts.
Optimised selection cuts for $M_H \gtrsim 600$ GeV are suggested.
}


\newcommand{\sla}[1]{\ifmmode%
  \setbox0=\hbox{$#1$}%
  \setbox1=\hbox to\wd0{\hss$/$\hss}\else%
  \setbox0=\hbox{#1}%
  \setbox1=\hbox to\wd0{\hss/\hss}\fi%
  #1\hskip-\wd0\box1 }
\newcommand{\calM}{{\cal M}}
\newcommand{\calO}{{\cal O}}

\begin{document}

% ===============================================================================

\maketitle

% ===============================================================================

\section{Introduction}

A key objective of particle physics research is the experimental 
confirmation of a theoretically consistent description of elementary 
particle masses and electroweak symmetry breaking.  The prevalent 
formalism is the Higgs mechanism 
\cite{Higgs:1964ia,Higgs:1964pj,Higgs:1966ev,Englert:1964et,Guralnik:1964eu}, 
which predicts the existence of one or more physical Higgs bosons.  
Recently, a candidate Standard Model (SM) Higgs boson with 
$M_H\approx 126$ GeV has been discovered \cite{Aad:2012tfa,Chatrchyan:2012ufa}. 
If physics beyond the SM (BSM) is realized in nature, 
additional, heavier Higgs bosons may be discovered at the 
Large Hadron Collider (LHC).  A comprehensive analysis of Higgs boson searches 
at the LHC can be found in 
refs.~\cite{Dittmaier:2011ti,Dittmaier:2012vm,Heinemeyer:2013tqa}.

At the LHC and Tevatron, Higgs bosons are primarily produced in 
gluon fusion \cite{Georgi:1977gs}.  
Next-to-leading order (NLO) QCD corrections have been calculated in the heavy-top 
limit \cite{Dawson:1990zj} and  with finite $t$ and $b$ mass effects \cite{Djouadi:1991tka,Graudenz:1992pv,Spira:1995rr}, and were found to be as large as
80--100\% at the LHC.  
This motivated the calculation of next-to-next-to-leading 
order (NNLO) QCD corrections \cite{Harlander:2002wh,Anastasiou:2002yz,Ravindran:2003um} 
enhanced by soft-gluon resummation at next-to-next-to-leading logarithmic 
level \cite{Catani:2003zt,deFlorian:2011xf} and beyond \cite{Moch:2005ky}.
At NNLO QCD, the residual scale uncertainty is of $\calO(10\%)$ 
for inclusive observables \cite{Dittmaier:2011ti}.
In addition to higher-order QCD corrections, electroweak corrections 
have been computed and found to be at the 1--5\% 
level \cite{Djouadi:1994ge,Aglietti:2004nj,Degrassi:2004mx,Actis:2008ug,Anastasiou:2008tj}.

The $H\to WW\to \ell\bar{\nu}\bar{\ell}\nu$ decay mode\footnote{%
Charged leptons are denoted by $\ell$.}
in gluon fusion plays an important role in Higgs searches at the LHC 
\cite{Dittmar:1996ss,Dittmaier:2012vm}, and fully differential 
NNLO QCD \cite{Anastasiou:2007mz,Grazzini:2008tf,Anastasiou:2008ik} and 
NLO electroweak \cite{Bredenstein:2006rh} corrections have 
been calculated and studied for this process.\footnote{%
A comprehensive NLO QCD analysis of the irreducible $WW$+0,1 jet background 
including squared quark-loop contributions has been performed in ref.\ \cite{Cascioli:2013gfa}.}

With inclusive NNLO signal 
scale uncertainties of $\calO(10\%)$, which can be further reduced by 
experimental selection cuts, it is 
important to study signal-background interference effects, because 
they can be of similar size if invariant Higgs masses above the 
weak-boson pair threshold contribute.  An accurate estimate of 
the magnitude of signal-background interference effects allows 
experimenters to decide if it has to be taken into 
account or may be treated as an additional uncertainty.
We note that interference effects at the few percent level if neglected 
in SM calculations could be wrongly identified as anomalous couplings.

Higgs-continuum interference in $gg\ (\to H)\to WW$ and $gg\ (\to H)\to ZZ$ 
has been studied for a light and heavy SM Higgs boson 
in refs.\ \cite{Glover:1988fe,Glover:1988rg,Seymour:1995qg,Binoth:2006mf,Accomando:2007xc,Campbell:2011cu,Kauer:2012ma,Passarino:2012ri,Kauer:2012hd,Campanario:2012bh,Bonvini:2013jha,Caola:2013yja}.  Signal-background interference for 
$gg\ (\to H)\to WW\to \ell\bar{\nu}\bar{\ell}'\nu'$ (different-flavour final state) 
and Higgs masses between 120 GeV and 600 GeV has been analysed at LO in ref.\ 
\cite{Campbell:2011cu} as well as refs.\ 
\cite{Kauer:2012hd,Binoth:2006mf,Accomando:2007xc,Kauer:2012ma}, and 
including (N)NLO corrections in soft-collinear approximation in ref.\ 
\cite{Bonvini:2013jha} for $M_H=600$ GeV.  Signal-background interference for 
$gg\ (\to H)\to ZZ\to \ell\bar{\ell}\nu'\bar{\nu}'$ has been investigated in 
ref.\ \cite{Kauer:2012hd} for $M_H=125$ GeV and $M_H=200$ GeV. 

For the same-flavour final state 
$\ell\bar{\nu}_\ell\bar{\ell}\nu_\ell$ considered here, 
in addition to Higgs-continuum interference one also has
interference between $WW$ and $ZZ$ intermediate states.
The $H\to WW/ZZ\to\ell\bar{\nu}_\ell\bar{\ell}\nu_\ell$ 
decay including interference was calculated at LO and NLO in 
ref.\ \cite{Bredenstein:2006rh,Dittmaier:2012vm,Heinemeyer:2013tqa}
and interference-induced deviations of 10\% for the shape of distributions 
have been found.\footnote{%
We note that substantial interference effects between direct and indirect Higgs 
boson decays to quarkonia have been identified in ref.\ \cite{Bodwin:2013gca}.}
For continuum production, the $WW/ZZ$ interference has been studied 
in $q\bar{q}$ scattering at LO in ref.\ \cite{Melia:2011tj}.  
No sizeable effects were found.
Here, we investigate both types of interference for the 
gluon scattering subprocess at loop-induced LO.

The paper is organised as follows: In section \ref{sec:calculation}, 
calculational details are discussed.  $WW/ZZ$ and signal-background 
interference is studied for the gluon-induced continuum background as well
as light and heavy Higgs boson signals with minimal and realistic experimental 
selection cuts in section \ref{sec:results}.
Conclusions are given in section \ref{sec:conclusions}.

% ===============================================================================

\section{Calculational details\label{sec:calculation}}

We implement the $gg\ (\to H) \to WW/ZZ\to\ell\bar{\nu}_\ell\bar{\ell}\nu_\ell$ 
process (same-flavour final state) in the publicly available parton-level program 
and event generator \textsf{gg2VV} \cite{gg2VV}, thus completing the implemention 
of all $gg\ (\to H) \to 4$ leptons processes at loop-induced leading order (LO).

% -------------------------------------------------------

\subsection{Amplitude\label{sec:process}}

\begin{figure}[t]
\vspace{0.4cm}
\centering
\includegraphics[width=0.8\textwidth, clip=true]{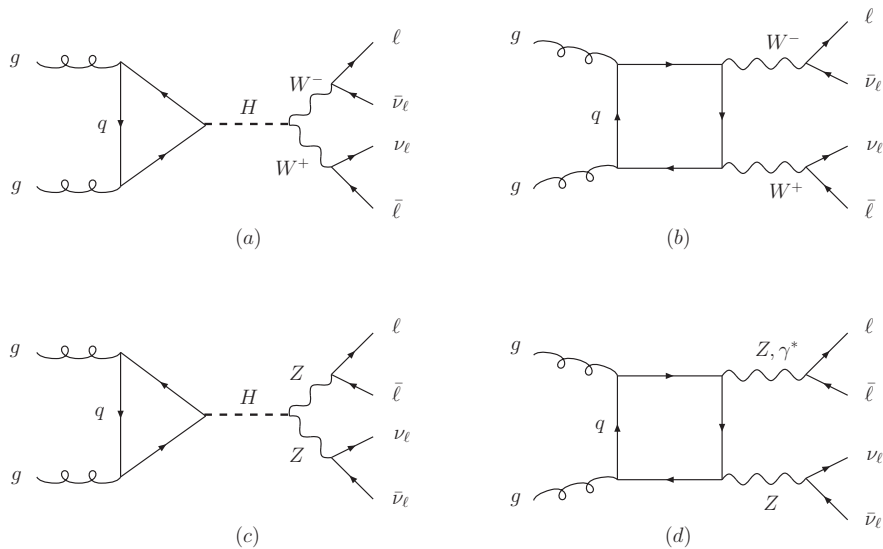}\\[.0cm]
\caption{\label{fig:graphs}
Representative Feynman graphs for the LO amplitude of $gg\to WW \to \ell\bar{\nu}_\ell\bar{\ell}\nu_\ell$ (Higgs signal (a) and continuum background (b) contributions)
and $gg\to ZZ \to \ell\bar{\nu}_\ell\bar{\ell}\nu_\ell$ (Higgs signal (c) and continuum background (d) contributions).}
\end{figure}

\begin{figure}[t]
\vspace{0.4cm}
\centering
\includegraphics[height=2.6cm, clip=true]{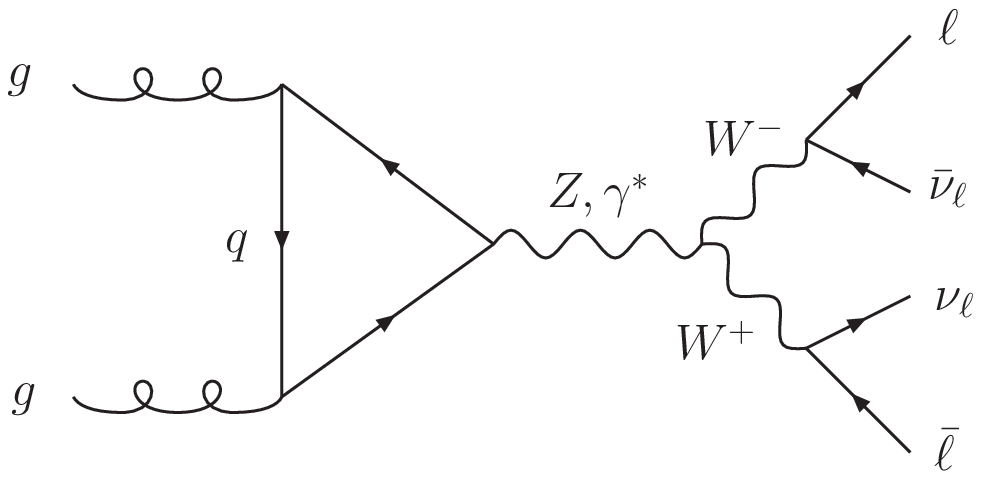}\hfil
\includegraphics[height=2.6cm, clip=true]{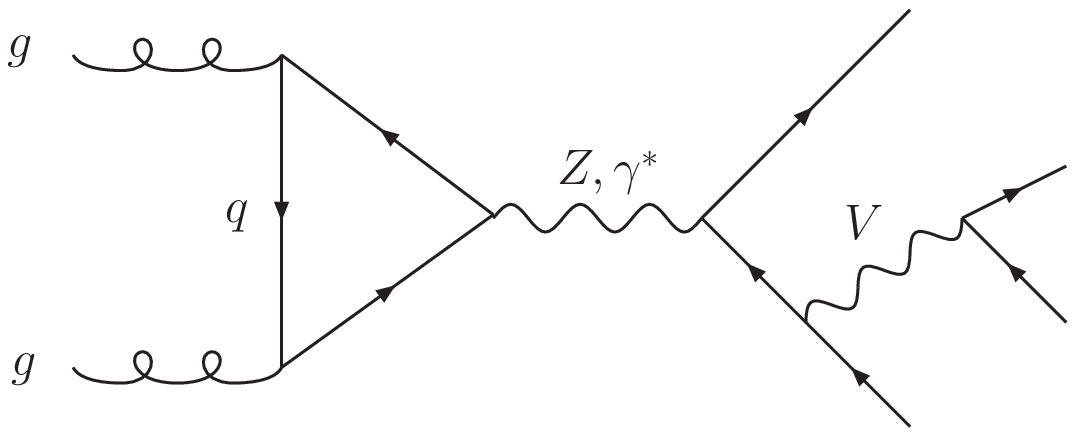}
\caption{\label{fig:triangles}
Representative triangle graphs that formally contribute to $gg\to \ell\bar{\nu}_\ell\bar{\ell}\nu_\ell$.}
\end{figure}

Figure \ref{fig:graphs} displays representative graphs for the Higgs signal 
process (a,c) and the $gg$-initiated continuum background process (b,d). 
For the same-flavour final state, $WW$ intermediate states (a,b) and 
$ZZ$ intermediate states (c,d) contribute.  Note that the 
amplitude contributions (a), (b), (c) and (d) interfere. 
$\gamma^\ast\to \ell\bar{\ell}$ contributions are important for 
Higgs (invariant) masses below the $Z$-pair threshold \cite{Binoth:2008pr} 
and are therefore included.
In addition to box topologies, in principle also triangle topologies 
contribute to the $gg$ continuum process (see figure \ref{fig:triangles}).
But, in the limit of vanishing lepton masses the triangle graphs do not contribute.\footnote{Note that the $gg\to Z^\ast$ triangle graphs do contribute for non-zero lepton masses, which was verified by explicit calculation.}

The implementation of the amplitudes for signal and background 
processes is generated with FeynArts/FormCalc \cite{Hahn:2000kx,Hahn:1998yk}
and subsequently customised.  The Higgs amplitudes are implemented using 
the complex-pole scheme described in ref.\ \cite{Goria:2011wa}.
The $gg$ continuum amplitude receives contributions from box graphs
that are affected by numerical instabilities when Gram determinants 
approach zero.  In these critical phase space regions the amplitude 
is evaluated in quadruple precision.  Residual instabilities are eliminated 
by requiring that $p_{T,W}$ and $p_{T,Z}$ are larger than $1$ GeV.

% -------------------------------------------------------

\subsection{Phase space\label{sec:sampling}}

We use a modified version of the adaptive-importance-sampling-oriented phase space integration method employed in 
ref.\ \cite{Melia:2011tj} for the same-flavour $ZZ\to \ell\bar{\ell}\ell\bar{\ell}$ 
decay mode.  Note that for a given final state, amplitudes with two distinct, 
potentially resonant intermediate states related via permutation contribute to the 
decay: $(Z\to \ell_1\bar{\ell}_1, Z\to \ell_2\bar{\ell}_2)$ and 
$(Z\to \ell_1\bar{\ell}_2, Z\to \ell_2\bar{\ell}_1)$.  For the different-flavour 
case, efficient importance sampling can be achieved by applying a mapping that 
approximates the Breit-Wigner resonance as well as the $\gamma^\ast$-induced 
divergence at zero virtuality when integrating over both (physically distinct) 
$Z$ boson virtualities $s_i= p_i^2$, where $p_i$ is the 4-momentum of the 
intermediate $Z$ boson state and $i=1,2$ is the flavour index.  To efficiently 
sample the same-flavour final state, ref.\ \cite{Melia:2011tj} uses an 
appropriately weighted sum of the two phase space integrals corresponding to the 
two lepton permutations $(\bar{\ell}_1\leftrightarrow\bar{\ell}_2)$.  The 
phase-space-configuration-dependent weight for each channel is chosen such that 
not efficiently sampled resonant configurations are suppressed.  Proper 
normalisation is achieved by overall 
multiplication with the inverse sum of the weights.\footnote{As noted in 
ref.\ \cite{Melia:2011tj}, in the $ZZ\to \ell\bar{\ell}\ell\bar{\ell}$ case it is 
advisable to randomly permute the identical leptons if final state criteria 
(e.g.\ selection cuts) are applied.}  When adapting the method to 
$(Z\to \ell\bar{\ell}, Z\to \nu_\ell\bar{\nu}_\ell)$ and 
$(W\to \ell\bar{\nu}_\ell, W\to \bar{\ell}\nu_\ell)$ an asymmetry is introduced, 
and to preserve efficiency it is necessary to numerically balance both weight 
factors.  This can be achieved by using
\begin{align}
f_{ZZ} &= \left\{\left[(s_{12}-M_W^2)^2 + (M_W\Gamma_W)^2\right] \times \left[(s_{34}-M_W^2)^2 + (M_W\Gamma_W)^2\right]\right\}^2\,,\\
f_{WW} &= \frac{s_{13}^3}{M_Z^6}\ \left[(s_{13}-M_Z^2)^2 + (M_Z\Gamma_Z)^2\right]^2 \times \frac{s_{24}^3}{M_Z^6}\ \left[(s_{24}-M_Z^2)^2 + (M_Z\Gamma_Z)^2\right]^2\,, 
\end{align}
with 4-momentum indices $1=\ell$, $2=\bar{\nu}_\ell$, $3=\bar{\ell}$, $4=\nu_\ell$.
The integrated cross section is computed using normalised weight factors:
\begin{gather}
\sigma = \int_{WW} \frac{f_{WW}}{f_{WW}+f_{ZZ}}\ d\sigma + \int_{ZZ}  \frac{f_{ZZ}}{f_{WW}+f_{ZZ}}\ d\sigma\,,
\end{gather}
where the integral subscript indicates which phase space mapping is applied.

Adaptive Monte Carlo integration was carried out using the Dvegas 
package \cite{Dvegas}, which was developed in the context of 
refs.\ \cite{Kauer:2001sp,Kauer:2002sn}.
The correctness of the program was checked with cross section and squared 
amplitude level comparisons with GoSam \cite{Cullen:2011ac}, MCFM \cite{Campbell:1999ah,Campbell:2011bn,Campbell:2011cu} and VBFNLO \cite{Arnold:2008rz,Arnold:2011wj}.

% ===============================================================================

\section{Results\label{sec:results}}

In this section, we present cross sections 
and measures for interference and off-shell effects for the 
$gg\to\ell\bar{\nu}_\ell\bar{\ell}\nu_\ell$ process 
in $pp$ collisions at $\sqrt{s}=8$ TeV.  The complete loop-induced
LO amplitude is included, i.e.\ Higgs signal and 
continuum background contributions as well as $WW$ and $Z(Z,\gamma^\ast)$
intermediate states (collectively denoted by $VV$).
All results are given for a single lepton flavour combination.  
No flavour summation is carried out for charged leptons or neutrinos.\footnote{%
Since our study focuses on interference effects, we do not include 
final states with different neutrino flavours, which cannot be distinguished 
experimentally.  Results for the corresponding different-flavour 
final states have been presented in ref.\ \cite{Kauer:2012hd} and 
can be computed with \textsf{gg2VV}.}
As input parameters, we use the specification of the 
LHC Higgs Cross Section Working Group in App.~A of ref.\ \cite{Dittmaier:2011ti}
with NLO weak boson widths and $G_\mu$ scheme.  
Finite top and bottom quark mass effects 
are included.  Lepton masses are neglected.
The fixed-width prescription is used for weak boson propagators.
The complex-pole-scheme Higgs widths are calculated with 
the program \textsf{HTO} (G.\ Passarino, unpublished).
We consider the Higgs masses 126 GeV, 200 GeV, 400 GeV, 600 GeV, 800 GeV and 1 TeV
with $\Gamma_H = 4.17116$ MeV, 1.42120 GeV, 26.5977 GeV, 103.933 GeV, 235.571 GeV 
and 416.119 GeV, respectively.
The PDF set MSTW2008 NNLO \cite{Martin:2009iq} with 3-loop running for 
$\alpha_s(\mu^2)$ and $\alpha_s(M_Z^2)=0.11707$ is used.
The CKM matrix is set to the unit matrix, which causes a 
negligible error \cite{Kauer:2012ma}.
Unless otherwise noted, the renormalisation and factorisation scales are 
fixed at $\mu_R=\mu_F=\mu_H= M_H/2$, which
yields a better perturbative convergence for the signal than
$\mu_R=\mu_F=M_H$ \cite{Anastasiou:2002yz}.
In section \ref{sec:lighthiggs}, the dynamic scale choice 
$\mu_R=\mu_F=\mu_\text{offshell}= M_{\ell\bar{\nu}_\ell\bar{\ell}\nu_\ell}/2$
is employed for comparison.
This choice was proposed in ref.\ \cite{Caola:2013yja} for
the far-off-shell region, $M_{\ell\bar{\nu}_\ell\bar{\ell}\nu_\ell} > 2M_Z$,
which gives a significant contribution to the total cross section
\cite{Kauer:2012hd}. We note that at Higgs resonance, one has $\mu_H\approx \mu_\text{offshell}$.  
When a full NLO calculation of the interference becomes available, the optimal
scale choice should be investigated in more detail.%
\footnote{(N)NLO corrections to signal-background interference in 
$gg\ (\to H) \to WW\to\ell\bar{\nu}\bar{\ell}'\nu'$ have been calculated in 
soft-collinear approximation in ref.\ \cite{Bonvini:2013jha}.}
When considering the continuum background only (section \ref{sec:ggcont}), 
the fixed scale $\mu_R=\mu_F= (M_W+M_Z)/2$ is used.

The resonant process $gg\to H \to\ell\bar{\nu}_\ell\bar{\ell}\nu_\ell$ 
(amplitude ${\cal M}_H$) and continuum process 
$gg\to\ell\bar{\nu}_\ell\bar{\ell}\nu_\ell$
(amplitude ${\cal M}_\text{cont}$) have the same initial and final states.
The amplitudes therefore interfere.  Furthermore, 
amplitude contributions with $W^-W^+$ and $ZZ$
intermediate states interfere.   By choosing suitable lepton flavour 
combinations, the $W^-W^+$ and $ZZ$ contributions can be separated.  
Both are therefore gauge invariant.  We denote the amplitudes 
corresponding to the graph sets (a), (b), (c) and (d) of 
figure \ref{fig:graphs} with ${\cal M}_{H,WW}$, ${\cal M}_{\text{cont},WW}$, 
${\cal M}_{H,ZZ}$ and ${\cal M}_{\text{cont},ZZ}$, respectively. 
The interference is given by
\begin{equation}
I_{ij} = 2\,\mbox{Re}({\cal M}_i{\cal M}^\ast_j) = 
|{\cal M}_i + {\cal M}_j|^2 - \rho_i - \rho_j \,,
\end{equation}
where $\rho_{i,j}= |{\cal M}_{i,j}|^2$.
As suggestive relative, symmetric, non-negative interference measure, we consider
\begin{equation}
R(i,j)=\frac{\sigma(\rho_i+\rho_j+I_{ij})}{\sigma(\rho_i+\rho_j)} \,,
\end{equation}
and apply it to Higgs signal and continuum background contributions 
($i=H$, $j=\text{cont}$) as well as contributions from $WW$ and $ZZ$ intermediate 
states ($i=WW$, $j=ZZ$):
\begin{align}
R_1 &= R(H,\text{cont}) \,, \\
R_3 &= R(WW,ZZ) \,. \label{eq:r3}
\end{align}
The contributions are described in section \ref{sec:process}.
When two interfering amplitude contributions are not viewed on an equal 
footing, the interference as relative correction to the primary contribution 
is a suggestive asymmetric measure:
\begin{equation}
\widetilde{R}(i,j)=\frac{\sigma(\rho_i+I_{ij})}{\sigma(\rho_i)} \,.
\end{equation}
Employing this definition, a Higgs-boson-search-inspired alternative to $R_1$ is 
given by 
\begin{equation}
R_2= \widetilde{R}(H,\text{cont}) \,.
\end{equation}
It measures the relative signal modification due to signal-background 
interference, because the null hypothesis, i.e.\ the SM without Higgs, 
is not sensitive to the interference.
In addition to $R_3$ given in eq.\ (\ref{eq:r3}), we also compute the $WW/ZZ$ interference as relative 
correction to the cross section of the dominant intermediate weak boson state,
either $WW$ or $ZZ$ depending on the applied (search) selection cuts:
\begin{equation}
R_4 = \widetilde{R}(WW,ZZ)\qquad\text{and}\qquad R_5 = \widetilde{R}(ZZ,WW) \,,
\end{equation}
respectively.
Finally, for comparison the relative correction of the gluon-induced cross section 
due to the subdominant intermediate weak boson state is given when interference is 
neglected:
\begin{equation}
R_6 = \sigma(\rho_{ZZ})/\sigma(\rho_{WW})\qquad\text{and}\qquad  
R_7 = \sigma(\rho_{WW})/\sigma(\rho_{ZZ}) \,,
\end{equation}
respectively.\footnote{%
Note that due to additional contributions from undetectable non-interfering  
neutrino flavour combinations the phenomenological corrections are $3R_6$
and $R_7/3$.}

% ===============================================================================

\subsection{Gluon-induced continuum background\label{sec:ggcont}}

First, we consider $WW/ZZ$ interference in the SM without Higgs boson.
Continuum weak-boson pair production in gluon scattering 
\cite{Glover:1988fe,Glover:1988rg,Dicus:1987dj,%
Zecher:1994kb,Binoth:2005ua,Binoth:2006mf,Binoth:2008pr,Campbell:2011bn,%
Campbell:2011cu,Frederix:2011ss} formally contributes to 
$pp\to \ell\bar{\nu}_\ell\bar{\ell}\nu_\ell$ at NNLO,\footnote{%
We note that electroweak corrections to $pp\to VV$ have been computed in 
refs.\ \cite{Bierweiler:2012kw,Bierweiler:2013dja,Baglio:2013toa} 
and threshold-resummed and approximate NNLO results for $pp\to WW$ in ref.\ \cite{Dawson:2013lya}.}
but is enhanced by the large gluon-gluon flux at the LHC (and is 
further enhanced by Higgs search selection cuts \cite{Binoth:2005ua}).\footnote{%
We further note that cross sections for gluon-induced $VV+$jet production have 
been calculated in refs.\ 
\cite{Melia:2012zg,Agrawal:2012df,Agrawal:2012as,Campanario:2012bh}.}
We note that $WW/ZZ$ interference in continuum production in quark-antiquark 
scattering was calculated at LO in ref.\ \cite{Melia:2011tj} and found to be 
negligible.
Here, results for the gluon-induced 
continuum process $gg\to \ell\bar{\nu}_\ell\bar{\ell}\nu_\ell$ are presented 
in table \ref{tab:ggcont}. 
In addition to a minimal $M_{\ell\bar{\ell}} > 4$ GeV cut,
we also consider LHC standard cuts 
for $WW$ production ($p_{T\ell} > 20$ GeV, $|\eta_\ell| < 2.5$, 
$\sla{p}_T > 45$ GeV, $M_{\ell\bar{\ell}} > 12$ GeV,  
$|M_{\ell\bar{\ell}} - M_Z| > 15$ GeV) and $ZZ$ production 
($p_{T\ell} > 20$ GeV, $|\eta_\ell| < 2.5$, 
$|M_{\ell\bar{\ell}} - M_Z| < 15$ GeV) \cite{Aad:2013wqa,Chatrchyan:2013lba}.
The renormalisation and factorisation scales are set to
$(M_W+M_Z)/2$.
\begin{table}[t]
\vspace{0.4cm}
\renewcommand{\arraystretch}{1.2}
\centering
{\footnotesize
\begin{tabular}{|l|ccc|cc|c|}
\cline{2-4} 
\multicolumn{1}{c|}{} & \multicolumn{3}{|c|}{$gg\ \to WW/ZZ \to \ell\bar{\nu}_\ell\bar{\ell}\nu_\ell$,} & \multicolumn{3}{|c}{} \\
\cline{5-7} 
\multicolumn{1}{c|}{} & \multicolumn{3}{|c|}{$\sigma$ [fb], $pp$, $\sqrt{s}=8$ TeV} & \multicolumn{2}{c|}{interference} & \multicolumn{1}{c|}{correction} \\
\hline
cuts & $WW$ & $ZZ$ & $WW/ZZ$ & $R_3$ & $R_{4/5}$ & $R_{6/7}$ \\
\hline
minimal       & 18.780(5) & 2.4581(6) & 21.241(4) & 1.0001(3) & 1.0002(4) & 0.13089(5) \\
standard $WW$ & 6.243(2) & 0.03914(2) & 6.285(2) & 1.0003(4) & 1.0003(4) & 0.006270(3) \\
standard $ZZ$ & 2.3815(7) & 1.5267(4) & 3.9116(8) & 1.0009(3) & 1.0023(8) & 1.5599(6) \\
\hline
\end{tabular}}\\[.0cm]
\caption{\label{tab:ggcont}
Cross sections for $gg\ \to WW/ZZ \to \ell\bar{\nu}_\ell\bar{\ell}\nu_\ell$ in $pp$ collisions at $\sqrt{s}=8$ TeV for three sets of selection cuts calculated at loop-induced leading order.
The interference measures  
$R_3=\sigma(|\calM_{WW}+\calM_{ZZ}|^2)/\sigma(|\calM_{WW}|^2+|\calM_{ZZ}|^2)$, $R_4=\sigma(|\calM_{WW}|^2+2\,\mbox{Re}({\cal M}_{WW}{\cal M}^\ast_{ZZ}))/\sigma(|\calM_{WW}|^2)$ and $R_5=\sigma(|\calM_{ZZ}|^2+2\,\mbox{Re}({\cal M}_{WW}{\cal M}^\ast_{ZZ}))/\sigma(|\calM_{ZZ}|^2)$ as well as the correction measures $R_6=\sigma(|\calM_{ZZ}|^2)/\sigma(|\calM_{WW}|^2)$ and $R_7=\sigma(|\calM_{WW}|^2)/\sigma(|\calM_{ZZ}|^2)$ are also displayed.  $R_{4/5} = R_4\ (R_5)$ and  $R_{6/7} = R_6\ (R_7)$ for minimal and standard $WW$ cuts (standard $ZZ$ cuts).  
Minimal cuts: $M_{\ell\bar{\ell}} > 4$ GeV.
$WW$ standard cuts: $p_{T\ell} > 20$ GeV, $|\eta_\ell| < 2.5$, 
$\sla{p}_T > 45$ GeV, $M_{\ell\bar{\ell}} > 12$ GeV and 
$|M_{\ell\bar{\ell}} - M_Z| > 15$ GeV.
$ZZ$ standard cuts: $p_{T\ell} > 20$ GeV, $|\eta_\ell| < 2.5$, 
$|M_{\ell\bar{\ell}} - M_Z| < 15$ GeV.
$\gamma^\ast$ contributions are included in $ZZ$.  
Cross sections are given for a single lepton flavour combination.  
The integration error is displayed in brackets.
}
\end{table}
The results demonstrate that $WW/ZZ$ interference effects in gluon-fusion 
continuum production are negligible.  Table \ref{tab:ggcont} also demonstrates 
that the $ZZ$ correction to $WW$ production is of ${\cal O}(10\%)$ with minimal 
selection cuts, but becomes negligible when standard $WW$ selection cuts are applied.
In contrast, the $WW$ correction to $ZZ$ production is of ${\cal O}(1)$ 
when standard $ZZ$ selection cuts are applied.

% ===============================================================================
% ===============================================================================
% ===============================================================================

\subsection{Light Higgs boson signal\label{sec:lighthiggs}}

We now study signal-background and $WW/ZZ$ interference effects 
in the Standard Model with a $126$ GeV Higgs boson motivated by 
the recent discovery \cite{Aad:2012tfa,Chatrchyan:2012ufa}.
Results for minimal, standard and Higgs search cuts 
are shown in tables \ref{tab:light_minimal}, \ref{tab:light_standard} 
and \ref{tab:light_higgssearch}, respectively.
As above, a minimal $M_{\ell\bar{\ell}} > 4$ GeV cut is applied
to exclude the on-shell photon singularity.
In this section, we apply the following cuts as standard cuts for the 
$WW \to \ell\bar{\nu}_\ell\bar{\ell}\nu_\ell$ process 
at the LHC:\footnote{The staggered $p_{T\ell}$ cut 
accomodates the off-shell $W$ boson decay for $M_{WW}\approx 126$ GeV.
Leptons are ordered by decreasing $p_T$.}
$p_{T\ell,\text{1st}} > 25$ GeV, $p_{T\ell,\text{2nd}} > 15$ GeV, $|\eta_\ell| < 2.5$, $\sla{p}_T > 45$ GeV, $M_{\ell\bar{\ell}} > 12$ GeV and 
$|M_{\ell\bar{\ell}} - M_Z| > 15$ GeV.
As $H\to WW$ search cuts, we apply these standard cuts and 
in addition $M_{\ell\bar{\ell}} <$ 50 GeV,  
$\Delta\phi_{\ell\bar{\ell}} < 1.8$ and $0.75\,M_H < M_{T} < M_H$ 
\cite{Aad:2013wqa}.\footnote{Very similar 
selection cuts are employed in ref.\ \cite{Chatrchyan:2013lba}.}  
The cut on the transverse mass
\begin{gather}
\label{eq:MT1}
M_{T1}=\sqrt{(M_{T,\ell\bar{\ell}}+\sla{p}_{T})^2-({\bf{p}}_{T,\ell\bar{\ell}}+{\sla{\bf{p}}}_T)^2}\quad\text{with}\quad M_{T,\ell\bar{\ell}}=\sqrt{p_{T,\ell\bar{\ell}}^2+M_{\ell\bar{\ell}}^2}
\end{gather}
strongly reduces the contribution from 
$M_{\ell\bar{\nu}_\ell\bar{\ell}\nu_\ell} > M_H$ \cite{Campbell:2011cu}.\footnote{
In the absence of additional final state particles,
the expression for $M_{T1}$ simplifies due to ${\sla{\bf{p}}}_T=-{\bf{p}}_{T,\ell\ell}$.
}
The renormalisation and factorisation scales are set to
$\mu_R=\mu_F= M_H/2$, except in table \ref{tab:light_minimal_dynscale}.

\begin{table}[t]
\vspace{0.4cm}
\renewcommand{\arraystretch}{1.2}
\centering
{\footnotesize
\begin{tabular}{|c|ccc|cc|}
\cline{2-4} 
\multicolumn{1}{c|}{} & \multicolumn{3}{|c|}{$gg\ (\to H)\to VV \to \ell\bar{\nu}_\ell\bar{\ell}\nu_\ell$,} & \multicolumn{2}{|c}{} \\
\multicolumn{1}{c|}{} & \multicolumn{3}{|c|}{$\sigma$ [fb], $pp$, $\sqrt{s}=8$ TeV,} & \multicolumn{2}{|c}{} \\
\cline{5-6} 
\multicolumn{1}{c|}{} & \multicolumn{3}{|c|}{$M_H=126$ GeV, min.\ cuts} & \multicolumn{2}{c|}{interference} \\
\hline
$VV$ & $H$ & cont & $|H$+cont$|^2$ & $R_1$ & $R_2$ \\
\hline
$WW$ & 17.593(4) & 20.978(5) & 36.480(8) & 0.9458(3) & 0.8811(6) \\
$ZZ$ & 0.9308(3) & 2.7543(7) & 3.436(2) & 0.9324(5) & 0.732(2) \\
$WW/ZZ$ & 17.729(3) & 23.742(5) & 39.125(8) & 0.9434(3) & 0.8677(6) \\
\hline
$R_3$ & 0.9571(3) & 1.0004(3) & 0.9802(3) &  \multicolumn{2}{|c}{}\\
$R_4$ & 0.9548(3) & 1.0004(4) & 0.9783(3) &  \multicolumn{2}{|c}{}\\
$R_6$ & 0.05291(2) & 0.13129(5) & 0.09419(5) &  \multicolumn{2}{|c}{}\\
\cline{1-4} 
\end{tabular}}\\[.0cm]
\caption{\label{tab:light_minimal}
Cross sections and interference measures for 
$gg\ (\to H)\to WW/ZZ \to \ell\bar{\nu}_\ell\bar{\ell}\nu_\ell$ in $pp$ collisions 
at $\sqrt{s}=8$ TeV.
The off-shell Higgs cross section at $M_H=126$ GeV, the gluon-induced continuum cross section and 
the sum including interference are given.  
The signal-background interference measures 
$R_1=\sigma(|{\cal M}_H + {\cal M}_\text{cont}|^2)/\sigma(|{\cal M}_H|^2 + |{\cal M}_\text{cont}|^2)$ 
and 
$R_2=\sigma(|{\cal M}_H|^2+2\,\mbox{Re}({\cal M}_H{\cal M}^\ast_\text{cont}))/\sigma(|{\cal M}_H|^2)$ 
as well as several $WW/ZZ$-related interference and correction measures 
(see table \protect\ref{tab:ggcont})
are also displayed.
A minimal $M_{\ell\bar{\ell}} > 4$ GeV cut is applied.
Other details as in table \protect\ref{tab:ggcont}.
}
\end{table}

\begin{table}[t]
\vspace{0.4cm}
\renewcommand{\arraystretch}{1.2}
\centering
{\footnotesize
\begin{tabular}{|c|ccc|cc|}
\cline{2-4} 
\multicolumn{1}{c|}{} & \multicolumn{3}{|c|}{$gg\ (\to H)\to VV \to \ell\bar{\nu}_\ell\bar{\ell}\nu_\ell$,} & \multicolumn{2}{|c}{} \\
\multicolumn{1}{c|}{} & \multicolumn{3}{|c|}{$\sigma$ [fb], $pp$, $\sqrt{s}=8$ TeV,} & \multicolumn{2}{|c}{} \\
\multicolumn{1}{c|}{} & \multicolumn{3}{|c|}{$M_H=126$ GeV, min.\ cuts,} & \multicolumn{2}{|c}{} \\
\cline{5-6} 
\multicolumn{1}{c|}{} & \multicolumn{3}{|c|}{$\mu_R=\mu_F= M_{\ell\bar{\nu}_\ell\bar{\ell}\nu_\ell}/2$} & \multicolumn{2}{c|}{interference} \\
\hline
$VV$ & $H$ & cont & $|H$+cont$|^2$ & $R_1$ & $R_2$ \\
\hline
$WW$ & 17.318(4) & 16.925(4) & 32.803(8) & 0.9580(3) & 0.9169(6) \\
$ZZ$ & 0.8822(2) & 2.1553(6) & 2.872(1) & 0.9455(4) & 0.813(2) \\
$WW/ZZ$ & 17.402(3) & 19.084(4) & 34.884(7) & 0.9561(3) & 0.9079(5) \\
\hline
$R_3$ & 0.9562(3) & 1.0002(3) & 0.9778(3) &  \multicolumn{2}{|c}{}\\
$R_4$ & 0.9540(3) & 1.0002(4) & 0.9759(4) &  \multicolumn{2}{|c}{}\\
$R_6$ & 0.05094(2) & 0.12735(5) & 0.08756(4) &  \multicolumn{2}{|c}{}\\
\cline{1-4} 
\end{tabular}}\\[.0cm]
\caption{\label{tab:light_minimal_dynscale}
Cross sections and interference measures for 
$gg\ (\to H)\to WW/ZZ \to \ell\bar{\nu}_\ell\bar{\ell}\nu_\ell$ in $pp$ collisions 
at $\sqrt{s}=8$ TeV and $M_H=126$ GeV.
The dynamic scale   
$\mu_R=\mu_F= M_{\ell\bar{\nu}_\ell\bar{\ell}\nu_\ell}/2$
is employed.  
Details as in table \protect\ref{tab:light_minimal}.
}
\end{table}

For the Higgs signal cross section, we find negative $WW/ZZ$ interference of 
approximately $5\%$, whereas no significant $WW/ZZ$ interference occurs 
for the continuum background in agreement with our findings in 
section \ref{sec:ggcont}.  This difference can be traced back to the fact that 
for $H\to VV$ with $M_V < M_H < 2M_V$ the most likely configuration features 
one  weak boson ($W$ as well as $Z$) that is (far) off-shell, whereas for the 
continuum background no such kinematic constraint exists.

Signal-background interference effects for minimal, standard and Higgs search cuts 
are compatible with what would be expected on the basis of different-flavour final 
state results \cite{Campbell:2011cu,Kauer:2012hd}.  With minimal and standard cuts, 
the signal-background interference of the subdominant $ZZ$ contribution 
when measured with $R_2$ is significantly larger than for the dominant $WW$ 
contribution.  Due to the suppression of the $ZZ$ contribution (high with standard 
and search cuts), the induced change of the overall signal-background interference 
is, however, at most at the one-percentage-point level.    As seen in table 
\ref{tab:light_higgssearch}, the application of a $M_{T1} < M_H$ cut suppresses 
the interference effects even stronger for the $ZZ$ contribution than for the 
$WW$ contribution.  

Results for $\mu_R=\mu_F= M_{\ell\bar{\nu}_\ell\bar{\ell}\nu_\ell}/2$ and minimal cut are displayed in table \ref{tab:light_minimal_dynscale}.  Signal-background interference effects are smaller compared to $\mu_R=\mu_F= M_H/2$, because the dynamic scale reduces 
the contribution of the region with strong interference, due to the smaller strong coupling at higher $VV$ invariant mass.  $R_2$ decreases by 3--8 percentage points
relative to table \ref{tab:light_minimal}.  Changes in the $WW/ZZ$ 
interference measures are at the sub-percentage point level.

\begin{table}[t]
\vspace{0.4cm}
\renewcommand{\arraystretch}{1.2}
\centering
{\footnotesize
\begin{tabular}{|c|ccc|cc|}
\cline{2-4} 
\multicolumn{1}{c|}{} & \multicolumn{3}{|c|}{$gg\ (\to H)\to VV \to \ell\bar{\nu}_\ell\bar{\ell}\nu_\ell$,} & \multicolumn{2}{|c}{} \\
\multicolumn{1}{c|}{} & \multicolumn{3}{|c|}{$\sigma$ [fb], $pp$, $\sqrt{s}=8$ TeV,} & \multicolumn{2}{|c}{} \\
\cline{5-6} 
\multicolumn{1}{c|}{} & \multicolumn{3}{|c|}{$M_H=126$ GeV, std.\ cuts} & \multicolumn{2}{c|}{interference} \\
\hline
$VV$ & $H$ & cont & $|H$+cont$|^2$ & $R_1$ & $R_2$ \\
\hline
$WW$ & 3.606(2) & 8.084(3) & 10.442(3) & 0.8932(3) & 0.654(1) \\
$ZZ$ & 0.009752(6) & 0.05099(3) & 0.05387(3) & 0.8868(7) & 0.295(5) \\
$WW/ZZ$ & 3.545(3) & 8.17(2) & 10.46(2) & 0.893(3) & 0.646(8) \\
\hline
$R_3$ & 0.9804(8) & 1.004(3) & 0.996(2) &  \multicolumn{2}{|c}{}\\
$R_4$ & 0.9803(8) & 1.004(3) & 0.996(2) &  \multicolumn{2}{|c}{}\\
$R_6$ & 0.002704(2) & 0.006308(5) & 0.005159(4) &  \multicolumn{2}{|c}{}\\
\cline{1-4} 
\end{tabular}}\\[.0cm]
\caption{\label{tab:light_standard}
Cross sections and interference measures for 
$gg\ (\to H)\to WW/ZZ \to \ell\bar{\nu}_\ell\bar{\ell}\nu_\ell$ in $pp$ collisions 
at $\sqrt{s}=8$ TeV and $M_H=126$ GeV.
$WW$ standard cuts for $M_{WW}\approx 126$ GeV are applied: 
$p_{T\ell,\text{1st}} > 25$ GeV, $p_{T\ell,\text{2nd}} > 15$ GeV, $|\eta_\ell| < 2.5$, $\sla{p}_T > 45$ GeV, $M_{\ell\bar{\ell}} > 12$ GeV and 
$|M_{\ell\bar{\ell}} - M_Z| > 15$ GeV.
Other details as in table \protect\ref{tab:light_minimal}.
}
\end{table}

\begin{table}[t]
\vspace{0.4cm}
\renewcommand{\arraystretch}{1.2}
\centering
{\footnotesize
\begin{tabular}{|c|ccc|cc|}
\cline{2-4} 
\multicolumn{1}{c|}{} & \multicolumn{3}{|c|}{$gg\ (\to H)\to VV \to \ell\bar{\nu}_\ell\bar{\ell}\nu_\ell$,} & \multicolumn{2}{|c}{} \\
\multicolumn{1}{c|}{} & \multicolumn{3}{|c|}{$\sigma$ [fb], $pp$, $\sqrt{s}=8$ TeV,} & \multicolumn{2}{|c}{} \\
\multicolumn{1}{c|}{} & \multicolumn{3}{|c|}{$M_H=126$ GeV,} & \multicolumn{2}{|c}{} \\
\cline{5-6} 
\multicolumn{1}{c|}{} & \multicolumn{3}{|c|}{Higgs search cuts} & \multicolumn{2}{c|}{interference} \\
\hline
$VV$ & $H$ & cont & $|H$+cont$|^2$ & $R_1$ & $R_2$ \\
\hline
$WW$ & 2.9303(7) & 0.7836(4) & 3.6649(8) & 0.9868(4) & 0.9833(4) \\
$ZZ$ & 0.004658(3) & 0.002851(2) & 0.007494(3) & 0.9979(6) & 0.9966(9) \\
$WW/ZZ$ & 2.8758(7) & 0.7864(4) & 3.6131(8) & 0.9866(3) & 0.9829(4) \\
\hline
$R_3$ & 0.9799(4) & 0.9999(8) & 0.9839(3) &  \multicolumn{2}{|c}{}\\
$R_4$ & 0.9798(4) & 0.9999(8) & 0.9838(3) &  \multicolumn{2}{|c}{}\\
$R_6$ & 0.0015898(9) & 0.003638(3) & 0.002045(1) &  \multicolumn{2}{|c}{}\\
\cline{1-4} 
\end{tabular}}\\[.0cm]
\caption{\label{tab:light_higgssearch}
Cross sections and interference measures for 
$gg\ (\to H)\to WW/ZZ \to \ell\bar{\nu}_\ell\bar{\ell}\nu_\ell$ in $pp$ collisions 
at $\sqrt{s}=8$ TeV and $M_H=126$ GeV.
Higgs search cuts are applied, i.e.\ $WW$ standard cuts (as in  
table \protect\ref{tab:light_standard}) and $M_{\ell\bar{\ell}} <$ 50 GeV,  
$\Delta\phi_{\ell\bar{\ell}} < 1.8$ and $0.75\,M_H < M_{T1} < M_H$.  
The transverse mass $M_{T1}$
is defined in eq.\ (\ref{eq:MT1}) in the main text.
Other details as in table \protect\ref{tab:light_minimal}.
}
\end{table}

% ===============================================================================

\subsection{Heavy Higgs boson signal\label{sec:heavyhiggs}}

We now study the signal-background and $WW/ZZ$ interference in the 
same-flavour decay mode for a SM-like Higgs boson with mass between 200 GeV 
and 1 TeV.  A priori, the search for a SM Higgs boson in the intermediate and 
heavy mass range is a crucial task.  It continues without premature conclusion 
about the exact nature of the discovered boson at 126 GeV.  
More specifically, one can search for a heavier SM Higgs boson with the advantage 
that assumptions about the realized SM extension are not required.  Given that the 
discovered 126 GeV boson appears to have all characteristics of the SM Higgs boson, 
an alternative, slightly better motivated approach is to search for a second, 
heavier Higgs boson.  A first 
proposal for a framework for the interpretation of the continuing LHC Higgs 
searches at masses other than 126 GeV has been set out in ref.\ 
\cite{Heinemeyer:2013tqa}.  These searches are being conducted in 
a model-independent way by suitably rescaling SM predictions in order to 
preserve unitarity cancellations at high energies as well as by considering 
specific BSM benchmark models. 
It is straightforward to apply the rescaling procedure 
described in ref.\ \cite{Heinemeyer:2013tqa} to the results 
presented in this section.  We note that the program \textsf{gg2VV} can be used 
to study Higgs-continuum interference and off-shell effects for BSM scenarios 
with a SM-like Higgs boson with rescaled $Hgg$, $HWW$ and 
$HZZ$ couplings.  The analysis of BSM benchmark models is beyond 
the scope of this paper.

In this context, the question arises if Higgs-Higgs interference effects 
are small.  The SM-like Higgs boson at 126 GeV appears to have the expected 
tiny width of approx.\ 4 MeV.\footnote{Bounding $\Gamma_H$ at $M_H\approx 126$ GeV 
with LHC and Tevatron data has been studied in refs.\ \cite{Barger:2012hv,Dobrescu:2012td,Ellis:2013lra,Djouadi:2013qya,CMS:yva,Martin:2012xc,Cheung:2013kla,Dixon:2013haa,Belanger:2013xza,Caola:2013yja}.} 
The overlap of the extremely narrow Breit-Wigner lineshape with the lineshape of a 
heavier Higgs boson with experimentally discriminable mass is insignificant. 
Higgs-Higgs interference effects are thus expected to be 
negligible provided that off-shell effects \cite{Kauer:2012hd,Kauer:2013cga} 
are suppressed by the search selection cuts.\footnote{%
We note that the role of interference effects in analysing the tensor structure 
of the $HZZ$ coupling has been studied in ref.\ \cite{Chen:2013waa}.}

Both approaches, a heavy SM or BSM Higgs boson search at the LHC (denoted by 
``SM'' and ``BSM'' below), have been 
pursued by ATLAS and CMS for Higgs masses up to 1 TeV.
In the following, we quote the experimental exclusion limits for a heavy Higgs 
boson with SM properties.
The $H\to ZZ\to \ell\bar{\ell}\ell\bar{\ell}$ (same or different flavour) and 
$H\to ZZ\to \ell\bar{\ell}\nu\bar{\nu}$ channels 
yield the strongest limits.\footnote{%
All exclusion limits are given at 95\% confidence level.}
The four-charged-lepton channel has been studied by ATLAS (SM) 
\cite{ATLAS:2013nma} % ATLAS-CONF-2013-013
and CMS (SM) 
\cite{CMS:xwa}.  % CMS-PAS-HIG-13-002
An exclusion region of 130--827 GeV is given in the latter work.
Secondly, the $H\to ZZ\to \ell\bar{\ell}\nu\bar{\nu}$ channel has been 
studied by CMS (SM and BSM) \cite{CMS-PAS-HIG-13-014}
and ATLAS (SM, $M_H < 600$ GeV, 7 TeV data)
\cite{Aad:2012ora}.  % arXiv:1205.6744
CMS and ATLAS exclude a mass in the ranges 248--930 GeV
and 319--558 GeV, respectively.
Furthermore, the $H\to WW\to \ell\bar{\nu}\bar{\ell}\nu$ channel has been analysed 
by ATLAS (SM) \cite{ATLAS-CONF-2013-067} and CMS (SM) 
\cite{CMS:bxa}, % CMS-PAS-HIG-13-003
excluding the mass ranges 260--642 GeV and 128-600 GeV, respectively.
The semileptonic decay modes $H\to WW\to \ell\nu jj$ 
\cite{Aad:2012me,CMS-PAS-HIG-13-008}  % arXiv:1206.6074, CMS-PAS-HIG-13-008
and $H\to ZZ\to \ell\bar{\ell} jj$
\cite{Aad:2012oxa}  % arXiv:1206.2443
have also been studied by ATLAS (SM) and CMS (SM and BSM).
A combined analysis of $H\to ZZ$ and $H\to WW$ channels in the 
mass range 145 GeV to 1 TeV has been carried out by CMS \cite{Chatrchyan:2013yoa}
and an update is in preparation \cite{JianWangTalk}.

% MINIMAL CUTS ------------------------------------------------------------------

\subsubsection{Minimal cuts}

\begin{table}[t]
\vspace{0.4cm}
\renewcommand{\arraystretch}{1.2}
\centering
{\footnotesize
\begin{tabular}{|c|c|ccc|cc|c|}
\cline{3-5} 
\multicolumn{2}{c|}{} & \multicolumn{3}{|c|}{$gg\ (\to H)\to VV \to \ell\bar{\nu}_\ell\bar{\ell}\nu_\ell$,} & \multicolumn{3}{|c}{} \\
\multicolumn{2}{c|}{} & \multicolumn{3}{|c|}{$\sigma$ [fb], $pp$, $\sqrt{s} = 8$ TeV,} & \multicolumn{3}{|c}{} \\
\cline{6-7}
\multicolumn{2}{c|}{} & \multicolumn{3}{|c|}{minimal cut} & \multicolumn{2}{c|}{interference} & \multicolumn{1}{|c}{}\\
\hline
\multicolumn{1}{|c|}{$M_H$ [GeV]} & $VV$ & $H$ & cont & $|H$+cont$|^2$ & $R_1$ & $R_2$ & $H$/cont \\
\hline
\multirow{7}{*}{200} & $WW$ & 23.662(7) & 17.781(5) & 41.677(9) & 1.0057(3) & 1.0099(5) & 1.3307(6) \\
 & $ZZ$ & 3.125(1) & 2.327(2) & 5.565(2) & 1.0208(4) & 1.0363(7) & 1.3431(8) \\
 & $WW/ZZ$ & 26.781(7) & 20.114(5) & 47.242(9) & 1.0074(3) & 1.0129(5) & 1.3315(5) \\
\cline{2-8} 
 & $R_3$ & 0.9998(4) & 1.0003(4) & 1.0000(3) &  \multicolumn{3}{|c}{}\\
 & $R_4$ & 0.9998(4) & 1.0003(4) & 1.0000(3) &  \multicolumn{3}{|c}{}\\
 & $R_6$ & 0.13208(6) & 0.13086(8) & 0.13354(5) &  \multicolumn{3}{|c}{}\\
\hline
\multirow{7}{*}{400} & $WW$ & 8.548(3) & 14.078(4) & 22.405(5) & 0.9902(3) & 0.9742(8) & 0.6072(3) \\
 & $ZZ$ & 1.5053(7) & 1.8298(7) & 3.318(1) & 0.9948(4) & 0.9886(9) & 0.8226(5) \\
 & $WW/ZZ$ & 10.053(3) & 15.912(4) & 25.727(5) & 0.9908(3) & 0.9764(7) & 0.6318(3) \\
\cline{2-8} 
 & $R_3$ & 1.0000(5) & 1.0002(4) & 1.0002(3) &  \multicolumn{3}{|c}{}\\
 & $R_4$ & 1.0000(5) & 1.0003(4) & 1.0002(3) &  \multicolumn{3}{|c}{}\\
 & $R_6$ & 0.1761(1) & 0.12998(6) & 0.14809(5) &  \multicolumn{3}{|c}{}\\
\hline
\multirow{7}{*}{600} & $WW$ & 1.430(3) & 12.354(4) & 14.142(5) & 1.0260(5) & 1.250(5) & 0.1157(2) \\
 & $ZZ$ & 0.2600(4) & 1.6014(7) & 1.9016(8) & 1.0216(6) & 1.154(5) & 0.1624(3) \\
 & $WW/ZZ$ & 1.690(3) & 13.959(4) & 16.047(5) & 1.0255(4) & 1.236(4) & 0.1211(2) \\
\cline{2-8} 
 & $R_3$ & 1.000(2) & 1.0003(4) & 1.0002(4) &  \multicolumn{3}{|c}{}\\
 & $R_4$ & 1.000(3) & 1.0003(5) & 1.0003(5) &  \multicolumn{3}{|c}{}\\
 & $R_6$ & 0.1819(4) & 0.12963(7) & 0.13447(7) &  \multicolumn{3}{|c}{}\\
\hline
\multirow{7}{*}{800} & $WW$ & 0.2857(7) & 11.286(3) & 11.806(3) & 1.0202(4) & 1.82(2) & 0.02531(6) \\
 & $ZZ$ & 0.0527(2) & 1.4602(6) & 1.5425(6) & 1.0195(6) & 1.56(2) & 0.03611(8) \\
 & $WW/ZZ$ & 0.3384(7) & 12.750(3) & 13.352(3) & 1.0201(4) & 1.78(2) & 0.02654(6) \\
\cline{2-8} 
 & $R_3$ & 1.000(3) & 1.0003(4) & 1.0003(4) &  \multicolumn{3}{|c}{}\\
 & $R_4$ & 1.000(4) & 1.0003(4) & 1.0003(4) &  \multicolumn{3}{|c}{}\\
 & $R_6$ & 0.1845(6) & 0.12938(6) & 0.13065(6) &  \multicolumn{3}{|c}{}\\
\hline
\multirow{7}{*}{1000} & $WW$ & 0.0759(4) & 10.547(3) & 10.760(3) & 1.0129(4) & 2.81(6) & 0.00720(3) \\
 & $ZZ$ & 0.01409(4) & 1.3616(6) & 1.3939(6) & 1.0132(6) & 2.29(6) & 0.01034(3) \\
 & $WW/ZZ$ & 0.0900(4) & 11.911(3) & 12.157(3) & 1.0129(4) & 2.73(5) & 0.00755(3) \\
\cline{2-8} 
 & $R_3$ & 1.000(5) & 1.0003(4) & 1.0003(4) &  \multicolumn{3}{|c}{}\\
 & $R_4$ & 1.000(6) & 1.0003(4) & 1.0003(4) &  \multicolumn{3}{|c}{}\\
 & $R_6$ & 0.1856(9) & 0.12910(7) & 0.12955(7) &  \multicolumn{3}{|c}{}\\
\cline{1-5} 
\end{tabular}}\\[.0cm]
\caption{\label{tab:heavy_minimal}
Cross sections and interference measures for 
$gg\ (\to H)\to WW/ZZ \to \ell\bar{\nu}_\ell\bar{\ell}\nu_\ell$ in $pp$ collisions 
at $\sqrt{s}=8$ TeV with a Higgs boson mass in the range 200 GeV to 1 TeV.
A minimal $M_{\ell\bar{\ell}} > 4$ GeV cut is applied.
Other details as in table \protect\ref{tab:light_minimal}.
}
\end{table}

As in section \ref{sec:lighthiggs}, 
the renormalisation and factorisation scales are set to
$\mu_R=\mu_F= M_H/2$.
Again, we first give results for 
a minimal $M_{\ell\bar{\ell}} > 4$ GeV cut.  To cover the heavy Higgs 
mass range of interest, results are shown in table \ref{tab:heavy_minimal} 
for a Higgs mass of 200 GeV, 400 GeV, 600 GeV, 800 GeV and 1 TeV.

In contrast to the light Higgs case, we find that $WW/ZZ$ interference for 
the Higgs signal is negligible for all considered heavy 
Higgs masses.  This further supports our conclusion in section \ref{sec:lighthiggs}, 
which is that non-negligible $WW/ZZ$ interference effects occur only if 
at least one 
weak boson of the pair is dominantly off-shell due to kinematic constraints.
This assertion is also confirmed by the results given in sections 
\ref{sec:heavyhiggszz} and \ref{sec:heavyhiggsww},
where heavy Higgs search selection cuts are applied.
With minimal cut the $ZZ$ correction is of ${\cal O}(10$--$20\%)$ 
and effects a change of the signal-background interference at the few 
percentage-point level.  For Higgs masses beyond 700 GeV, signal-over-background 
ratios are at the percent and per mil level, and 
signal-background interference ($R_2-1$) is of ${\cal O}(1)$.

% H -> ZZ SEARCH CUTS -------------------------------------------------------

\subsubsection{$H\to ZZ$ search cuts\label{sec:heavyhiggszz}}

While the four-charged-lepton channel has the highest sensitivity in the 
$H\to ZZ$ search for masses below 500 GeV, the 
$H\to ZZ\to \ell\bar{\ell}\nu\bar{\nu}$ channel considered here dominates above 
500 GeV \cite{JianWangTalk}.
As above, we consider the Higgs mass range 200 GeV--1 TeV and note that  
for the lower end of this range the $H\to WW\to \ell\bar{\ell}\nu\bar{\nu}$
process can contribute as much as 70\% to the signal after selection cuts
\cite{Aad:2012ora}.  For our calculations, we adopt the analysis strategy 
of the recent study presented in ref.\ \cite{CMS-PAS-HIG-13-014}.
In more detail, as $H\to ZZ$ search cuts we apply the basic selection cuts 
$\ p_{T\ell} > 20$ GeV, $|\eta_\ell| < 2.5$ and 
$|M_{\ell\bar{\ell}} - M_Z| < 15$ GeV.
In addition, to select the signal contribution with $M_{ZZ}\approx M_H$ 
Higgs-mass-dependent bounds are imposed on 
the transverse mass $M_{T2}$ and $\sla{p}_T$.\footnote{%
In our calculation, $p_{T,\ell\bar{\ell}}$ is constrained by the 
$\sla{p}_T$ cut.} 
$M_{T2}$ is given by
\begin{gather}
\label{eq:MT2}
M_{T2}=\sqrt{ \left(M_{T,\ell\bar{\ell}}+\sla{M}_{T}\right)^2-({\bf{p}}_{T,\ell\bar{\ell}}+{\sla{\bf{p}}}_T)^2 }\quad\text{with}\quad 
\sla{M}_{T}=\sqrt{\sla{p}_{T}^2+M_{\ell\bar{\ell}}^2} \,.
\end{gather}
$M_{T,\ell\bar{\ell}}$ is defined in eq.\ (\ref{eq:MT1}).\footnote{%
In the absence of additional final state particles: 
$M_{T2}=2M_{T,\ell\bar{\ell}}=2\sla{M}_{T}$.} 
The employed $M_{T2}$ and $\sla{p}_T$ bounds are given in table \ref{tab:zzcuts}.
Note that an upper bound is imposed on $M_{T2}$ for all Higgs masses
in order to suppress increasingly large interference effects at high invariant 
masses.
\begin{table}[t]
\vspace{0.4cm}
\renewcommand{\arraystretch}{1.2}
\centering
{\footnotesize
\begin{tabular}{|l|ccccc|}
\hline
$M_H$ [GeV] &                   200 & 400 & 600 & 800 & 1000 \\
\hline
$M_{T2}$ lower bound [GeV] &    180 & 300 & 420 & 500 & 550 \\
$M_{T2}$ upper bound [GeV] &    220 & 450 & 700  & 900 & 1100 \\
$\sla{p}_T$ lower bound [GeV] &  20 &  90 & 110 & 150 & 170 \\
\hline
\end{tabular}}\\[.0cm]
\caption{\label{tab:zzcuts}
Higgs-mass-dependent $H\to ZZ$ search cuts.
Bounding cuts are applied to the transverse 
mass $M_{T2}$ defined in eq.\ (\ref{eq:MT2}) in the main text
and the missing transverse momentum $\sla{p}_T$.
}
\end{table}

\begin{table}[t]
\vspace{0.4cm}
\renewcommand{\arraystretch}{1.2}
\centering
{\footnotesize
\begin{tabular}{|c|c|ccc|cc|c|}
\cline{3-5} 
\multicolumn{2}{c|}{} & \multicolumn{3}{|c|}{$gg\ (\to H)\to VV \to \ell\bar{\ell}\nu_\ell\bar{\nu}_\ell$,} & \multicolumn{3}{|c}{} \\
\multicolumn{2}{c|}{} & \multicolumn{3}{|c|}{$\sigma$ [fb], $pp$, $\sqrt{s} = 8$ TeV,} & \multicolumn{3}{|c}{} \\
\cline{6-7}
\multicolumn{2}{c|}{} & \multicolumn{3}{|c|}{$H\to ZZ$ cuts} & \multicolumn{2}{c|}{interference} & \multicolumn{1}{|c}{} \\
\hline
\multicolumn{1}{|c|}{$M_H$ [GeV]} & $VV$ & $H$ & cont & $|H$+cont$|^2$ & $R_1$ & $R_2$ & $H$/cont \\
\hline
\multirow{7}{*}{200} & $ZZ$ & 1.7507(5) & 0.6224(3) & 2.5153(6) & 1.0600(4) & 1.0813(5) & 2.813(2) \\
 & $WW$ & 2.2529(9) & 1.0064(6) & 3.338(2) & 1.0243(5) & 1.0351(7) & 2.239(2) \\
 & $ZZ/WW$ & 3.997(1) & 1.6302(6) & 5.848(2) & 1.0392(3) & 1.0552(5) & 2.452(2) \\
\cline{2-8} 
 & $R_3$ & 0.9983(4) & 1.0009(6) & 0.9990(3) &  \multicolumn{3}{|c}{}\\
 & $R_5$ & 0.9961(8) & 1.002(2) & 0.9976(7) &  \multicolumn{3}{|c}{}\\
 & $R_7$ & 1.2869(6) & 1.617(2) & 1.3272(6) &  \multicolumn{3}{|c}{}\\
\hline
\multirow{7}{*}{400} & $ZZ$ & 0.8919(3) & 0.07674(7) & 0.9508(3) & 0.9816(5) & 0.9800(5) & 11.62(1) \\
 & $WW$ & 0.02340(2) & 0.01400(3) & 0.03909(4) & 1.045(2) & 1.073(2) & 1.672(4) \\
 & $ZZ/WW$ & 0.9154(3) & 0.09074(7) & 0.9901(3) & 0.9840(4) & 0.9824(5) & 10.088(9) \\
\cline{2-8} 
 & $R_3$ & 1.0001(5) & 1.000(2) & 1.0001(5) &  \multicolumn{3}{|c}{}\\
 & $R_5$ & 1.0001(5) & 1.000(2) & 1.0001(5) &  \multicolumn{3}{|c}{}\\
 & $R_7$ & 0.02623(2) & 0.1824(4) & 0.04111(4) &  \multicolumn{3}{|c}{}\\
\hline
\multirow{7}{*}{600} & $ZZ$ & 0.15889(9) & 0.02187(1) & 0.18949(9) & 1.0483(7) & 1.0549(8) & 7.263(5) \\
 & $WW/10^{-4}$ & $0.232(2)$ & 1.57(3) & 1.98(4) & 1.10(3) & 1.8(2) & 0.148(3) \\
 & $ZZ/WW$ & 0.15891(9) & 0.02204(1) & 0.18970(9) & 1.0484(7) & 1.0551(8) & 7.211(5) \\
\cline{2-8} 
 & $R_3$ & 1.0000(8) & 1.0002(7) & 1.0001(7) &  \multicolumn{3}{|c}{}\\
 & $R_5$ & 1.0000(8) & 1.0002(7) & 1.0001(7) &  \multicolumn{3}{|c}{}\\
 & $R_7$ & 0.000146(2) & 0.0072(2) & 0.00105(2) &  \multicolumn{3}{|c}{}\\
\hline
\multirow{7}{*}{800} & $ZZ$ & 0.03191(2) & 0.010964(4) & 0.05113(2) & 1.1928(6) & 1.2590(8) & 2.910(2) \\
 & $WW/10^{-5}$ & 0.069(3) & 2.91(9) & 3.1(1) & 1.04(5) & 3(2) & 0.024(2) \\
 & $ZZ/WW$ & 0.03191(2) & 0.010994(4) & 0.05117(2) & 1.1927(6) & 1.2591(8) & 2.902(2) \\
\cline{2-8} 
 & $R_3$ & 1.0000(7) & 1.0001(6) & 1.0000(5) &  \multicolumn{3}{|c}{}\\
 & $R_5$ & 1.0000(7) & 1.0001(6) & 1.0000(5) &  \multicolumn{3}{|c}{}\\
 & $R_7$ & $2.2(1)\!\cdot\! 10^{-5}$ & 0.00265(9) & 0.00061(2) &  \multicolumn{3}{|c}{}\\
\hline
\multirow{7}{*}{1000} & $ZZ$ & 0.008415(3) & 0.007268(2) & 0.020719(6) & 1.3211(5) & 1.5984(9) & 1.1578(5) \\
 & $WW/10^{-5}$ & 0.0086(4) & 1.22(5) & 1.25(5) & 1.02(6) & 4(8) & 0.0071(5) \\
 & $ZZ/WW$ & 0.008416(3) & 0.007281(2) & 0.020733(6) & 1.3208(5) & 1.5984(9) & 1.1558(5) \\
\cline{2-8} 
 & $R_3$ & 1.0000(5) & 1.0001(4) & 1.0000(4) &  \multicolumn{3}{|c}{}\\
 & $R_5$ & 1.0000(5) & 1.0001(4) & 1.0000(4) &  \multicolumn{3}{|c}{}\\
 & $R_7$ & 0.00001(1) & 0.00168(7) & 0.00060(3) &  \multicolumn{3}{|c}{}\\
\cline{1-5} 
\end{tabular}}\\[.0cm]
\caption{\label{tab:heavy_HtoZZ}
Cross sections and interference measures for 
$gg\ (\to H)\to ZZ/WW \to \ell\bar{\ell}\nu_\ell\bar{\nu}_\ell$ in $pp$ collisions 
at $\sqrt{s}=8$ TeV with a Higgs boson mass in the range 200 GeV to 1 TeV.
$H\to ZZ$ search cuts are applied: 
$p_{T\ell} > 20$ GeV, $|\eta_\ell| < 2.5$, $|M_{\ell\bar{\ell}} - M_Z| < 15$ GeV
and the $M_H$-dependent cuts displayed in table \protect\ref{tab:zzcuts}.
Other details as in table \protect\ref{tab:light_minimal}.
}
\end{table}

Applying these $H\to ZZ$ search cuts, we obtain the results shown 
in table \ref{tab:heavy_HtoZZ}, which demonstrate the absence of 
$WW/ZZ$ interference for the Higgs signal and continuum background.
Signal-background interference ($R_2-1$) ranges from $-2\%$ to 
$+60\%$ and is not affected by the $WW$ correction for $M_H\gtrsim 400$ GeV.

% H -> WW SEARCH CUTS -----------------------------------------------------------

\subsubsection{$H\to WW$ search cuts\label{sec:heavyhiggsww}}

We adopt the $H\to WW$ search cuts of ref.\ \cite{CMS:bxa}.\footnote{In ref.\ \cite{ATLAS-CONF-2013-067} the same-lepton-flavour final states have not been included.}
To simulate detector coverage, select $WW\to \ell\bar{\nu}\bar{\ell}\nu$ 
and suppress 
reducible backgrounds the following cuts are applied: 
$|\eta_\ell| < 2.5$, $\sla{p}_T > 20$ GeV, $M_{\ell\bar{\ell}} > 12$ GeV, 
$|M_{\ell\bar{\ell}} - M_Z| > 15$ GeV and 
$p_{T,\ell\bar{\ell}} > 45$ GeV.
To select the $H\to WW$ signal, following ref.\ \cite{CMS:bxa}, we 
define the transverse mass 
\begin{gather}
\label{eq:MT3}
M_{T3}= \sqrt{2\,p_{T,\ell\bar{\ell}}\:\sla{p}_T(1-\cos\Delta\phi_{\ell\bar{\ell},\text{miss}})}\,,
\end{gather}
where $\Delta\phi_{\ell\bar{\ell},\text{miss}}$ is the angle between
${\bf{p}}_{T,\ell\bar{\ell}}$ and ${\sla{\bf{p}}}_T$,
and impose the cuts\ \,
$p_{T\ell,\text{max}} > 0.2\,M_H$, 
$p_{T\ell,\text{min}} > 25$ GeV, 
$M_{\ell\bar{\ell}} < 90$ GeV for $M_H=200$ GeV and  
$M_{\ell\bar{\ell}} < M_H - 100$ GeV for $M_H\geq 400$ GeV, 
$\Delta\phi_{\ell\bar{\ell}} < 100^\circ$ for $M_H=200$ GeV and 
$\Delta\phi_{\ell\bar{\ell}} < 175^\circ$ for $M_H\geq 400$ GeV, 
and 120 GeV $< M_{T3} < M_H$.\footnote{In the absence of additional final 
state particles: $M_{T3}=2\,p_{T,\ell\bar{\ell}}=2\,\sla{p}_T$.}

\begin{table}[t]
\vspace{0.4cm}
\renewcommand{\arraystretch}{1.2}
\centering
{\footnotesize
\begin{tabular}{|c|c|ccc|cc|c|}
\cline{3-5} 
\multicolumn{2}{c|}{} & \multicolumn{3}{|c|}{$gg\ (\to H)\to VV \to \ell\bar{\nu}_\ell\bar{\ell}\nu_\ell$,} & \multicolumn{3}{|c}{} \\
\multicolumn{2}{c|}{} & \multicolumn{3}{|c|}{$\sigma$ [fb], $pp$, $\sqrt{s} = 8$ TeV,} & \multicolumn{3}{|c}{} \\
\cline{6-7}
\multicolumn{2}{c|}{} & \multicolumn{3}{|c|}{$H\to WW$ cuts} & \multicolumn{2}{c|}{interference} & \multicolumn{1}{|c}{} \\
\hline
\multicolumn{1}{|c|}{$M_H$ [GeV]} & $VV$ & $H$ & cont & $|H$+cont$|^2$ & $R_1$ & $R_2$ & $H$/cont \\
\hline
\multirow{7}{*}{200} & $WW$ & 3.686(2) & 1.5060(8) & 5.614(2) & 1.0812(4) & 1.1144(6) & 2.448(2) \\
 & $ZZ$ & 0.008678(7) & 0.003983(3) & 0.012141(8) & 0.9589(8) & 0.940(2) & 2.179(3) \\
 & $WW/ZZ$ & 3.701(2) & 1.5100(8) & 5.633(2) & 1.0809(4) & 1.1139(6) & 2.451(2) \\
\cline{2-8} 
 & $R_3$ & 1.0017(5) & 1.0000(7) & 1.0012(4) &  \multicolumn{3}{|c}{}\\
 & $R_4$ & 1.0017(5) & 1.0000(7) & 1.0012(4) &  \multicolumn{3}{|c}{}\\
 & $R_6$ & 0.002354(2) & 0.002645(3) & 0.002163(2) &  \multicolumn{3}{|c}{}\\
\hline
\multirow{7}{*}{400} & $WW$ & 2.5009(9) & 0.6197(3) & 2.9806(9) & 0.9551(4) & 0.9440(5) & 4.036(3) \\
 & $ZZ$ & 0.03929(3) & 0.00983(2) & 0.04896(4) & 0.997(1) & 0.996(2) & 3.995(9) \\
 & $WW/ZZ$ & 2.5401(9) & 0.6298(4) & 3.0297(9) & 0.9558(4) & 0.9448(5) & 4.033(3) \\
\cline{2-8} 
 & $R_3$ & 1.0000(5) & 1.0004(7) & 1.0000(5) &  \multicolumn{3}{|c}{}\\
 & $R_4$ & 1.0000(5) & 1.0004(7) & 1.0000(5) &  \multicolumn{3}{|c}{}\\
 & $R_6$ & 0.01571(2) & 0.01587(4) & 0.01643(2) &  \multicolumn{3}{|c}{}\\
\hline
\multirow{7}{*}{600} & $WW$ & 0.5260(3) & 0.2659(2) & 0.8386(4) & 1.0590(6) & 1.0888(9) & 1.978(2) \\
 & $ZZ$ & 0.006632(5) & 0.00347(2) & 0.01099(2) & 1.087(3) & 1.133(4) & 1.909(8) \\
 & $WW/ZZ$ & 0.5326(3) & 0.2694(2) & 0.8496(4) & 1.0593(6) & 1.0893(9) & 1.977(2) \\
\cline{2-8} 
 & $R_3$ & 1.0000(7) & 1.0002(7) & 1.0000(6) &  \multicolumn{3}{|c}{}\\
 & $R_4$ & 1.0000(7) & 1.0002(7) & 1.0000(6) &  \multicolumn{3}{|c}{}\\
 & $R_6$ & 0.01261(2) & 0.01306(6) & 0.01310(3) &  \multicolumn{3}{|c}{}\\
\hline
\multirow{7}{*}{800} & $WW$ & 0.10346(5) & 0.10562(5) & 0.23894(9) & 1.1428(6) & 1.289(2) & 0.9795(7) \\
 & $ZZ$ & 0.0012318(8) & 0.001252(9) & 0.00291(1) & 1.173(6) & 1.35(2) & 0.984(7) \\
 & $WW/ZZ$ & 0.10469(5) & 0.10689(5) & 0.24185(9) & 1.1431(6) & 1.289(2) & 0.9794(7) \\
\cline{2-8} 
 & $R_3$ & 1.0000(7) & 1.0001(7) & 1.0000(6) &  \multicolumn{3}{|c}{}\\
 & $R_4$ & 1.0000(7) & 1.0001(7) & 1.0000(6) &  \multicolumn{3}{|c}{}\\
 & $R_6$ & 0.01191(1) & 0.01186(9) & 0.01219(4) &  \multicolumn{3}{|c}{}\\
\hline
\multirow{7}{*}{1000} & $WW$ & 0.02426(2) & 0.04357(2) & 0.08049(3) & 1.1867(6) & 1.522(2) & 0.5568(4) \\
 & $ZZ$ & 0.0002958(2) & 0.000512(4) & 0.000986(4) & 1.220(7) & 1.60(2) & 0.577(4) \\
 & $WW/ZZ$ & 0.02455(2) & 0.04408(2) & 0.08148(3) & 1.1871(6) & 1.523(2) & 0.5570(4) \\
\cline{2-8} 
 & $R_3$ & 1.0000(7) & 1.0001(7) & 1.0000(6) &  \multicolumn{3}{|c}{}\\
 & $R_4$ & 1.0000(7) & 1.0001(7) & 1.0000(6) &  \multicolumn{3}{|c}{}\\
 & $R_6$ & 0.01219(1) & 0.01176(8) & 0.01225(5) &  \multicolumn{3}{|c}{}\\
\cline{1-5} 
\end{tabular}}\\[.0cm]
\caption{\label{tab:heavy_HtoWW}
Cross sections and interference measures for 
$gg\ (\to H)\to WW/ZZ \to \ell\bar{\nu}_\ell\bar{\ell}\nu_\ell$ in $pp$ collisions 
at $\sqrt{s}=8$ TeV with a Higgs boson mass in the range 200 GeV to 1 TeV.
$H\to WW$ search cuts are applied: 
$p_{T\ell,\text{max}} > 0.2\,M_H$, 
$p_{T\ell,\text{min}} > 25$ GeV, $|\eta_\ell| < 2.5$, 
$\sla{p}_T > 20$ GeV, $p_{T,\ell\bar{\ell}} > 45$ GeV, 
$M_{\ell\bar{\ell}} > 12$ GeV, 
$|M_{\ell\bar{\ell}} - M_Z| > 15$ GeV, 
$M_{\ell\bar{\ell}} < 90$ GeV for $M_H=200$ GeV and  
$M_{\ell\bar{\ell}} < M_H - 100$ GeV for $M_H\geq 400$ GeV, 
$\Delta\phi_{\ell\bar{\ell}} < 100^\circ$ for $M_H=200$ GeV and 
$\Delta\phi_{\ell\bar{\ell}} < 175^\circ$ for $M_H\geq 400$ GeV, 
120 GeV $< M_{T3} < M_H$.
The transverse mass $M_{T3}$
is defined in eq.\ (\ref{eq:MT3}) in the main text.
Other details as in table \protect\ref{tab:light_minimal}.
}
\end{table}

With these $H\to WW$ search cuts, we obtain the results shown 
in table \ref{tab:heavy_HtoWW}.  The cuts reduce the $ZZ$ correction 
to the one-percent level, and the signal-background interference 
is dominated by the $WW$ contribution.

% ===============================================================================

\section{Conclusions\label{sec:conclusions}}

We studied $WW/ZZ$ interference for Higgs signal and continuum background as well 
as signal-background interference 
for same-flavour $\ell\bar{\nu}_\ell\bar{\ell}\nu_\ell$ final 
states produced in gluon-gluon scattering at the LHC for light and heavy 
Higgs masses with minimal and realistic experimental selection cuts.
For the signal cross section, we find 
$WW/ZZ$ interference effects of ${\cal O}(5\%)$ at $M_H=126$ GeV.  
For $M_H\ge 200$ GeV, we find that $WW/ZZ$ interference is negligible.  
For the $gg$ continuum background, we also find that $WW/ZZ$ interference 
is negligible.  
As general rule, we conclude that non-negligible $WW/ZZ$ interference 
effects occur only if at least one weak boson of the pair is dominantly 
off-shell due to kinematic constraints.
Our results demonstrate that in cases where $WW/ZZ$ interference can be 
neglected it is nevertheless important to take into account the subdominant
weak boson pair as its contribution is of ${\cal O}(10$--$20\%)$ before 
search selection cuts are applied.  Considering $WW$ intermediate 
states in $H\to ZZ$ searches is crucial due to the intrinsically larger 
continuum cross section.
The subdominant weak boson pair contribution induces a 
correction to the signal-background interference, which is 
at the few percentage point level except when selection cuts 
strongly suppress the subdominant contribution.
To mitigate the signal reduction in heavy Higgs searches due to increasingly large 
negative signal-background interference for invariant masses above the Higgs 
mass, we employed an upper bound cut on the transverse mass also for 
$M_H \gtrsim 600$ GeV.
The kinematical dependence of signal-background and $WW/ZZ$ interference 
for $gg\ \to H \to WW/ZZ\to\ell\bar{\nu}_\ell\bar{\ell}\nu_\ell$ 
can be calculated and simulated with the public parton-level 
event generator \textsf{gg2VV}.

% ===============================================================================

\acknowledgments

I would like to thank the conveners and 
members of the LHC Higgs Cross Section Working Group, 
notably S.\ Bolognesi, D.\ de Florian, S.\ Diglio, C.\ Mariotti, 
G.\ Passarino and R.\ Tanaka, for stimulating 
and informative discussions,
G.\ Heinrich, M.\ Rauch and M.\ Rodgers for useful comparisons, 
and the Centre for Particle Physics at Royal Holloway, University of London 
and the Institute for Theoretical 
Particle Physics and Cosmology at RWTH Aachen University for access to 
their computing facilities.
Financial support from the Higher Education Funding Council for England, 
the Science and Technology Facilities Council and the Institute for 
Particle Physics Phenomenology, Durham, is gratefully acknowledged.

% ===============================================================================

% ===============================================================================

\end{document}